\documentclass[runningheads]{llncs}
\usepackage{amsmath,amsfonts,amssymb}
\usepackage{bm}
\usepackage{wrapfig}
\usepackage{graphicx}
\usepackage{verbatim}
\begin{document}
\title{Contrast Adaptive Tissue Classification
	by Alternating Segmentation and Synthesis
}
\titlerunning{Alternating Segmentation and Synthesis}
%

\author{Dzung L. Pham\inst{1,2,3} \and
Yi-Yu Chou\inst{1,3} \and
Blake E. Dewey\inst{2} \and 
Daniel S. Reich\inst{3} \and ~~~~~
John A. Butman\inst{3} \and 
Snehashis Roy\inst{3}}
\authorrunning{D.L. Pham et al.}
%
\institute{Henry M. Jackson Foundation, Bethesda, MD 20892, USA \email{\{dzung.pham,yiyu.chou\}@nih.gov}\and
Johns Hopkins University, Baltimore, MD 21218, USA \email{blake.dewey@jhu.edu}\and
National Institutes of Health, Bethesda, MD 20892, USA \email{reichds@ninds.nih.gov,jbutmana@cc.nih.gov,snehashis.roy@nih.gov}}

%
\maketitle              
\begin{abstract}
Deep learning approaches to the segmentation of magnetic resonance images have shown significant promise in automating the quantitative analysis of brain images.  However, a continuing challenge  has been its sensitivity to the variability of acquisition protocols.  Attempting to segment images that have different contrast properties from those within the training data generally leads to significantly reduced performance.  Furthermore, heterogeneous data sets cannot be easily evaluated because the quantitative variation due to acquisition differences often dwarfs the variation due to the biological differences that one seeks to measure.  In this work, we describe an approach using alternating segmentation and synthesis steps that adapts the contrast properties of the training data to the input image.  This allows input images that do not resemble the training data to be more consistently segmented.  A notable advantage of this approach is that only a single example of the acquisition protocol is required to adapt to its contrast properties.  We demonstrate the efficacy of our approaching using brain images from a set of human subjects scanned with two different T1-weighted volumetric protocols.

\keywords{segmentation, synthesis, magnetic resonance imaging, harmonization, domain adaptation.}
\end{abstract}
\section{Introduction}
Automated segmentation algorithms for quantifying brain structure in magnetic resonance (MR)
images are widely used in neuroscientific research, and increasingly being applied in clinical trials and diagnostic applications.  The best performing algorithms currently rely on
training data or atlases to serve as exemplars of how MR images should be segmented.  However, MR image contrast is notoriously sensitive to both hardware differences (e.g., scanner manufacturers, receiver coils) and software differences (e.g., pulse sequence parameters, software platform versions). Segmentation approaches requiring training data perform suboptimally when faced with imaging data that possess contrast properties that differ from the atlases.  Perhaps more importantly, these algorithms provide inconsistent results when applied to heterogeneously acquired imaging data~\cite{Glocker2019,Shinohara17}.   Techniques for generating more consistent segmentations across different acquisition protocols are therefore needed to enable more accurate monitoring of anatomical changes, as well as for increasing statistical power~\cite{George2020}. 

Multiple approaches have been previously proposed to perform harmonization of heterogeneous imaging data.  A common approach is to incorporate site effects in the statistical modeling~\cite{Chua15,Fortin16}.  Such approaches are designed for group analyses and are complementary to image-based harmonization methods. Intensity normalization techniques that attempt to align the histograms of images using linear or piecewise linear transformations have also been proposed~\cite{nyul2000new,Shah11}.  Because these transformations affect the global histogram, local contrast differences and differences in the overall anatomy across images are not well addressed.  Image synthesis techniques, where a set of images from a subject is
used in combination with training data to create a new image with desirable intensity properties, have also been used for harmonization~\cite{Jog15}.  A recent approach used subjects scanned with  multiple acquisition protocols to serve as training data for the synthesis~\cite{Dewey19}.  Such training data is often not available, however.  

An alternative to harmonizing the appearances of images is to use a segmentation algorithm that is robust to variations in pulse sequences.  Generative and multi-atlas segmentation algorithms have been proposed~\cite{Puonti16,Erus18}, as well as deep learning algorithms employing domain adaptation~\cite{Kamnitsas17,Dou2019,Chen2020}.  The latter approaches typically employ  adversarial learning that are capable of addressing contrast variations without the need for paired training data.  A disadvantage of domain adaptation approaches is that they require multiple examples of the different acquisition protocols to adequately learn features invariant to the different domains.  A deep learning segmentation approach that does not require a groups of training data is described in~\cite{Jog19} that employs a parametric model of MRI image formation to augment the appearance of the training data.  

In this work, the appearance of the atlas MR images is altered to resemble the contrast of the input image without changing the atlas labels.  This is accomplished by using alternating steps of segmentation and synthesis.  A preliminary demonstration of this framework using Gaussian classifiers and a synthesis based on a linear combination of tissue memberships was previously described in~\cite{Pham18}.  The approach here employs deep learning networks for both the segmentation and synthesis processes, and more extensive evaluation results are shown to demonstrates its efficacy.  

\begin{figure}[t]
	\centering
	\includegraphics[height=4cm]{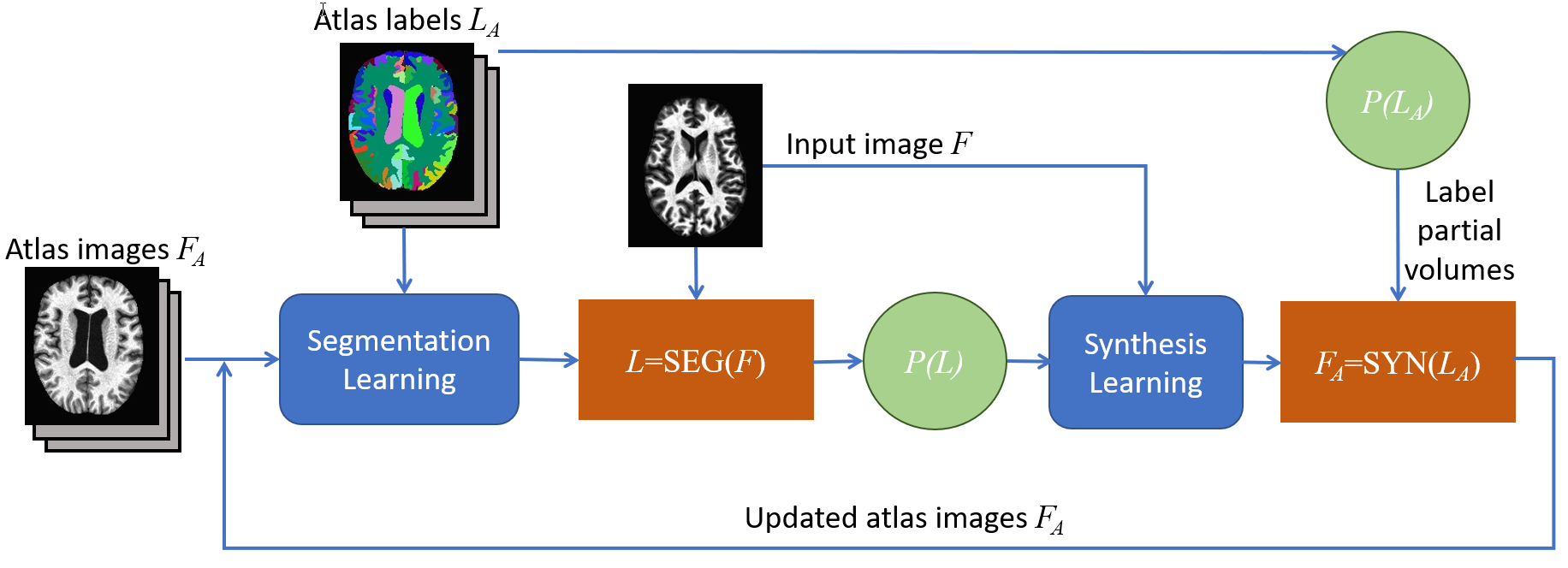} 
	\caption{\footnotesize Block diagram depicting the CAMELION framework.  An intermediate segmentation of the input data is used to train a synthesis network, mapping atlas labels to a contrast similar to the input. Blue blocks are training, red are testing, and green are partial volume estimation.}
	\vspace{-0.1in}
	\label{fig:block}
\end{figure}

\section{Methods}

We denote a segmentation $\mathrm{SEG}(\cdot)$ to be a mapping from an input MR image $F$ to its anatomical labels $L$, which represent the desired regions of interest.  We consider supervised segmentations that require a training data set consisting of MR image and label image pairs, $F_A$ and $L_A$. We denote the synthesis $\mathrm{SYN}(\cdot)$ to be a mapping from the labels to an MR
image. Our framework, referred to as CAMELION (Contrast Adaptive Method for Label Identification), uses these dual operations to update both the desired segmentation $L$ and the appearance of the atlas images $F_A$, while keeping the input image $F$ and the atlas labels $L_A$ fixed.  Fig.~\ref{fig:block} illustrates this process.  In the first iteration, the segmentation is performed as usual with the available training data, yielding an initial segmentation estimate $L^{(0)}$.  Because of differences between the atlas MR images $F_{A}^{(0)}$ and the input image $F$, this segmentation will be suboptimal.  To compensate, the initial segmentation and input MR image are used to train a synthesis network, which is then applied to the atlas label images to generate new atlas MR images $F_{A}^{(1)}$.  The segmentation network is retrained with the updated atlas images, and a new segmentation of the input MRI image, $L^{(1)}$, is computed.  This process is repeated until convergence.  

An advantage of the CAMELION framework is that only a single example of the input image's acquisition protocol is required to train the synthesis.  If multiple input images with the same protocol are available, the processing could be performed group-wise, potentially with better results because of increased training data.  In this work, however, we focus on processing each input image data set independently without pre-training any of the networks.

\vspace{-0.06in}
\subsection{Data}
\vspace{-0.06in}

\begin{wrapfigure}{r}{0.3\textwidth}
\vspace{-0.45in}
\begin{center}
	\begin{minipage}{0.95\textwidth}
		\tabcolsep 1pt
		\begin{tabular}[ht]{cc}
			\includegraphics[height=2cm]{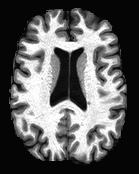} & 
			\includegraphics[height=2cm]{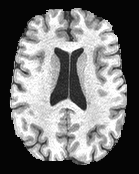}\\[-1ex]
			\scriptsize MPRAGE & \scriptsize SPGR \\
		\end{tabular}
	\end{minipage}
	\vspace{-0.2in}
	\caption{\footnotesize T1-w images from the same subject}
	\label{fig:data}
\end{center}
\vspace{-0.4in}
\end{wrapfigure} 
Data for this study was collected under an IRB-approved protocol from 18 healthy subjects with  familial relation to a person with multiple sclerosis.  Subjects underwent MRI scanning with two different T1-weighted volumetric protocols on a Siemens Skyra 3T Scanner.  The first scan was a Magnetization Prepared Rapid Gradient Echo (MPRAGE) protocol acquired at 1mm isotropic spatial resolution (TR=3000ms, TE=3.03ms, TI=900ms, FA=$9^\circ$).  The second scan was a Spoiled Gradient Recalled (SPGR) protocol also acquired at 1mm isotropic resolution (TR=7.8ms, TE=3ms, FA=$18^\circ$).  Although both scans are T1-weighted, the contrast properties are quite different.  Both scans were rigidly co-registered~\cite{Avants14} and then processed to remove intensity inhomogeneities~\cite{Tustison10} and extracerebral tissue~\cite{Iglesias11}.  Fig.~\ref{fig:data} shows examples of pre-processed data from these two acquisitions.  

\vspace{-0.06in}
\subsection{Segmentation}
\vspace{-0.06in}

\begin{figure}[t]
	\begin{center}
		\tabcolsep 2pt
		\begin{tabular}[ht]{ccccc}
			\includegraphics[height=2.3cm]{figs/atlas_mprage.png}& 
			\includegraphics[height=2.3cm]{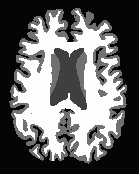}&
			\includegraphics[height=2.3cm]{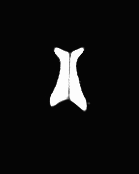}&
			\includegraphics[height=2.3cm]{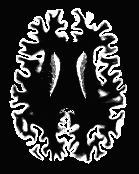}&
			\includegraphics[height=2.3cm]{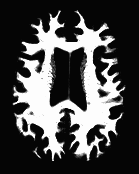} \\
			\footnotesize (a) & \footnotesize (b) & \footnotesize (c) & \footnotesize (d) &  \footnotesize (e) 
			\\
		\end{tabular}
	\end{center}
	\vspace{-0.2in}
	\caption{\footnotesize Axial slices from an atlas data set: (a) MPRAGE image, (b) MPRAGE segmentation using FreeSurfer, (c) ventricle partial volume, (d) gray matter partial volume, (e) white matter partial volume.  Partial volume images were computed from the MPRAGE and FreeSurfer segmentation and are used for image synthesis.}
	\vspace{-0.1in}
	\label{fig:atlas}
\end{figure}

The proposed framework is relatively agnostic to the specific segmentation algorithm, but does require that it sufficiently parcellates the brain to enable a reasonable synthesis.  In this work, we used a 3D U-net architecture~\cite{Cicek16} with five non-background tissue classes: cerebrospinal fluid (CSF), ventricles, gray matter, white matter, and brainstem.  The network  used three layers, a patch size of $64\times64\times64$, a batch size of 32, and a mean squared error loss function.  The Adam optimizer was used with a learning rate of 0.001.  In order to reduce training time, the network weights were saved between each iteration. A total of 30 epochs for the first iteration was used, and 15 epochs for successive iterations. Training data for the segmentation network was derived from  FreeSurfer~\cite{Fischl12} applied to MPRAGE images from ten subjects, with FreeSurfer labels merged appropriately into the five aforementioned tissue classes (see Fig.~\ref{fig:atlas}(b)).  White matter is shown in white, gray matter in gray, and CSF in dark gray.  Brainstem is not present in this particular slice.  Note that the CSF class in FreeSurfer does not include sulcal CSF, so was excluded from the evaluation results.

\vspace{-0.06in}
\subsection{Partial volume estimation}
\vspace{-0.06in}

Although it is possible to synthesize an image directly from discrete tissue labels, we have empirically found that synthetic images were more realistic using continuous labels, such as spatial probability functions or partial volume estimates.  The advantage of continuous labels is further increased because the synthesis network is trained with a single MR image and segmentation pair.  We therefore apply a partial volume estimation step following segmentation to generate a continuous function from zero to one for tissue class.  For each voxel $j$, the input image intensity $f_j$ is modeled as 
\begin{equation}
f_{j} = \sum_{k=1}^{K}p_{jk}c_k + \eta_j
\end{equation}
where $p_{jk}$ is the partial volume of class $k$ at $j$, $c_k$ is the mean intensity of tissue class $k$, and $\eta_j$ is a Gaussian noise term.  
In addition to the constraints that $p_{jk}>0$ and $\sum_{k=1}^{K} p_{jk} = 1$, we restrict partial volumes to only be greater than zero for a maximum of two classes. The two non-zero classes are determined by the associated discrete segmentation of the image as the tissue classes at the voxel and the spatially nearest tissue class different from the original class.  The mean intensity for each is directly estimated from the discrete segmentation and input image.

We further impose a prior probability distribution on $p_{jk}$ such that:
\begin{equation}
p_{jk} =
\begin{cases}
\frac{1}{Z}\exp \left({\beta(p_{jk}-0.5)^2}\right) & p_{jk} \in [0,1]\\
0 & \text{otherwise}
\end{cases}
\end{equation}
where $Z$ is a normalizing constant, and $\beta$ is a weighting parameter.  This prior allows the model to favor pure tissue classes over partial volume tissues for positive values of $\beta$.  A maximum a posteriori estimate of $p_{jk}$ can be straightforwardly computed for every $j$ using this model under the provided constraints. In this work, $\beta$ was determined empirically based on the visual quality of the partial volumes and then held fixed for all experiments.  Figs.~\ref{fig:atlas}(c)-(e) show an example result computed from the MPRAGE image in Fig.~\ref{fig:atlas}(a) and its segmentation in Fig.~\ref{fig:atlas}(b).

\vspace{-0.06in}
\subsection{Synthesis}
\vspace{-0.06in}

The synthesis process in this framework maps a set of tissue partial volumes back to the MR image.  Generating an MR image from a tissue classification has traditionally been performed in MR simulation approaches using physics-based models~\cite{Collins98}.  In our approach, the MR image formation process is implicitly modeled using a convolutional neural network. Given the input MR image $F$ and partial volumes $P(L)$, we optimize a 3D U-net with a mean squared error loss function defined by 
\begin{equation}
\min_\theta \|F - \hat{F}(P(L);\theta))\|_2^2
\end{equation}
where $\theta$ are the network weights.  U-net parameters were set similarly to the segmentation network, except the number of epochs was set to 20.  Once the synthesis is learned, new atlas MR images are generated from the atlas label images within the training data. The segmentation network is updated with this new training data, and the input image is once again segmented to compute a new segmentation estimate.

\section{Results}

We applied the U-net segmentation network trained with MPRAGE images and CAMELION to the SPGR images of 8 held out subjects.  For additonal comparison, we also applied the nonlinear histogram matching (NHM) method of~\cite{nyul2000new}.  To evaluate consistency, the segmentation resulting from applying the original U-net segmentation network to the corresponding MPRAGE image of the subject was used as the ground truth reference.  Convergence within CAMELION was empirically set to 5 iterations, which typically resulted in fewer than 5\% of voxels changing labels.  Note that each of the test images was processed completely independently, with the synthesis step trained using only the input image.  

\begin{figure}[t]
	\centering
	\begin{minipage}{0.95\textwidth}
		\tabcolsep 2pt
		\begin{tabular}[ht]{cccccc}
			\includegraphics[height=2.2cm]{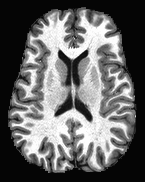} & 
			\includegraphics[height=2.2cm]{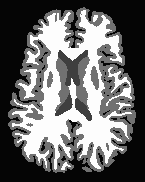} & 			
			\includegraphics[height=2.2cm]{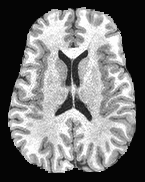}&
			\includegraphics[height=2.2cm]{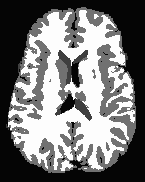}&
			\includegraphics[height=2.2cm]{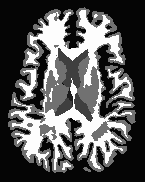}&
			\includegraphics[height=2.2cm]{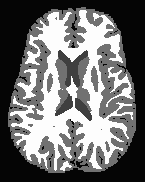} \\
			\footnotesize (a) & \footnotesize (b) & \footnotesize (c) & \footnotesize (d) & \footnotesize (e) &  \footnotesize (f)\\
		\end{tabular}
	\end{minipage}
	\vspace{-0.1in}
	\caption{\footnotesize Segmentation results: (a) MPRAGE image, (b) MPRAGE segmented with U-net, (c) SPGR image, (d) SPGR segmented with U-net, (e) SPGR segmented with U-net after nonlinear histogram matching, (f) SPGR segmented with CAMELION.}
	\vspace{-0.2in}
	\label{fig:test}
\end{figure}

Fig.~\ref{fig:test} shows the results of different segmentation approaches on the SPGR data at a midventricular axial slice.  Figs.~\ref{fig:test}(a)-(b) show the MPRAGE image and the U-net segmentation result that serves as the reference. Because the U-net segmentation is trained with MPRAGE data acquired with the same protocol, the segmentation result is reasonably accurate.  Figs.~\ref{fig:test}(c)-(d) show the associated SPGR image from the same subject, as well as the segmentation when applying the same U-net.  Because the SPGR contrast differs from the MPRAGE atlas images used in the training data, the white matter is overestimated and the gray matter is underestimated.  Fig.~\ref{fig:test}(e) shows the U-net segmentation applied to the SPGR after it has been intensity corrected using NHM.  Although the cortical gray matter is better estimated in this result, the subcortical gray matter is over estimated.  Fig.~\ref{fig:test}(f) shows the CAMELION result applied to the SPGR images.  By adapting the appearance of the atlas image to the SPGR input, the result is much more consistent with what would be obtained with an MPRAGE input.  Note that there are some differences at the cortical gray matter and sulcal CSF boundaries.  This is at least in part because the FreeSurfer algorithm does not explicitly segment sulcal CSF, leading to some ambiguity in the segmentation learning in these regions.

\begin{figure}[t]
	\centering
	\includegraphics[height=3.3cm]{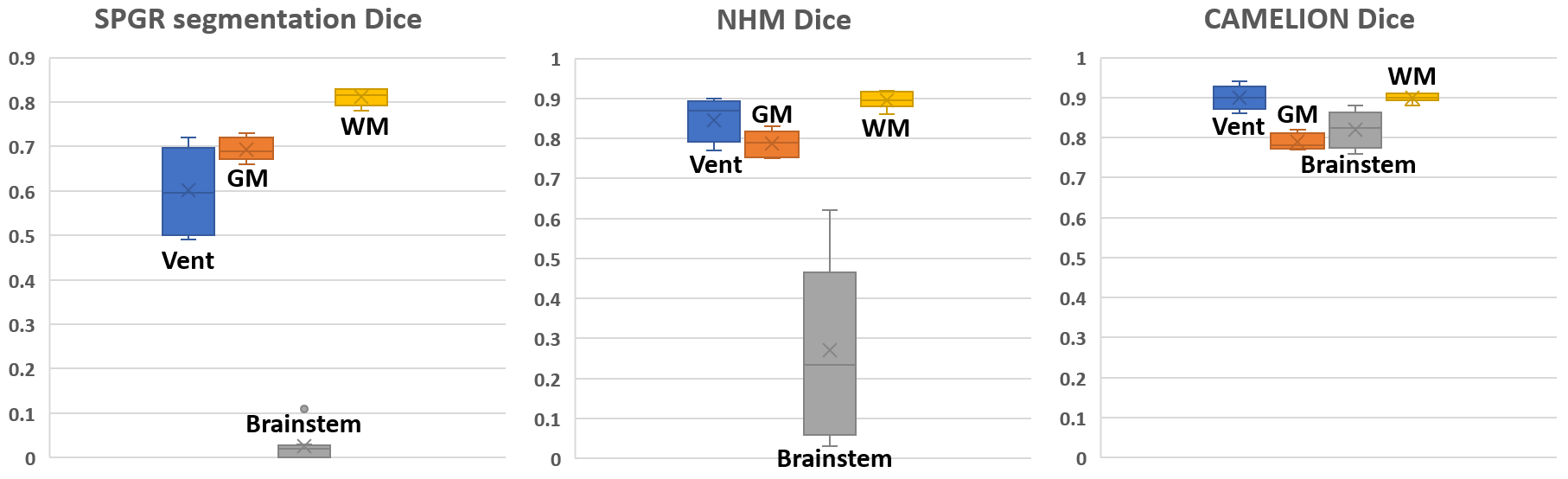} 
	\vspace{-0.1in}
	\caption{\footnotesize Quantitative box plot comparison of Dice coefficients using three different  approaches.  Four tissue classes were considered: ventricles (Vent), gray matter (GM), white matter (WM), and brainstem.  The center line is the median, the "x" is the mean, the box represents the first quartile, and the whiskers represent the minimum and maximum values.}
	\vspace{-0.1in}
	\label{fig:graph}
\end{figure}

Fig.~\ref{fig:graph} shows quantitative results of segmenting the SPGR test images, focusing on Dice overlap in  four tissue classes. Applying the U-net segmentation directly to the SPGR image yielded the lowest Dice coefficients.  Adjusting the global contrast of the SPGR images with NHM generally improved results, but smaller regions such as the brainstem remain poorly segmented.  Applying CAMELION further increased Dice coefficients and reduced variability of the measurements.  Improvements were statistically significant relative to the standard U-net segmentation across all four tissue classes, and were significant for ventricles and brainstem relative to the NHM results ($p<0.01$, paired t-test).
Although Dice overlap provides one measure of segmentation accuracy, in research studies the primary outcome is often the total volume of a structure.  Table~1 shows the Pearson correlation coefficient of volume measurements between the MPRAGE and SPGR segmentations of the same subject.  Bold values show the highest correlation across methods for that tissue class.  CAMELION (indicated by ``CAM'' in Table 1) shows improved associations between the volume measurements across the two acquisition protocols.

\begin{figure}[t]
	\begin{center}
	\begin{minipage}{0.95\textwidth}
	\tabcolsep 2pt
	\begin{tabular}[ht]{cccc}
		\includegraphics[height=2.5cm]{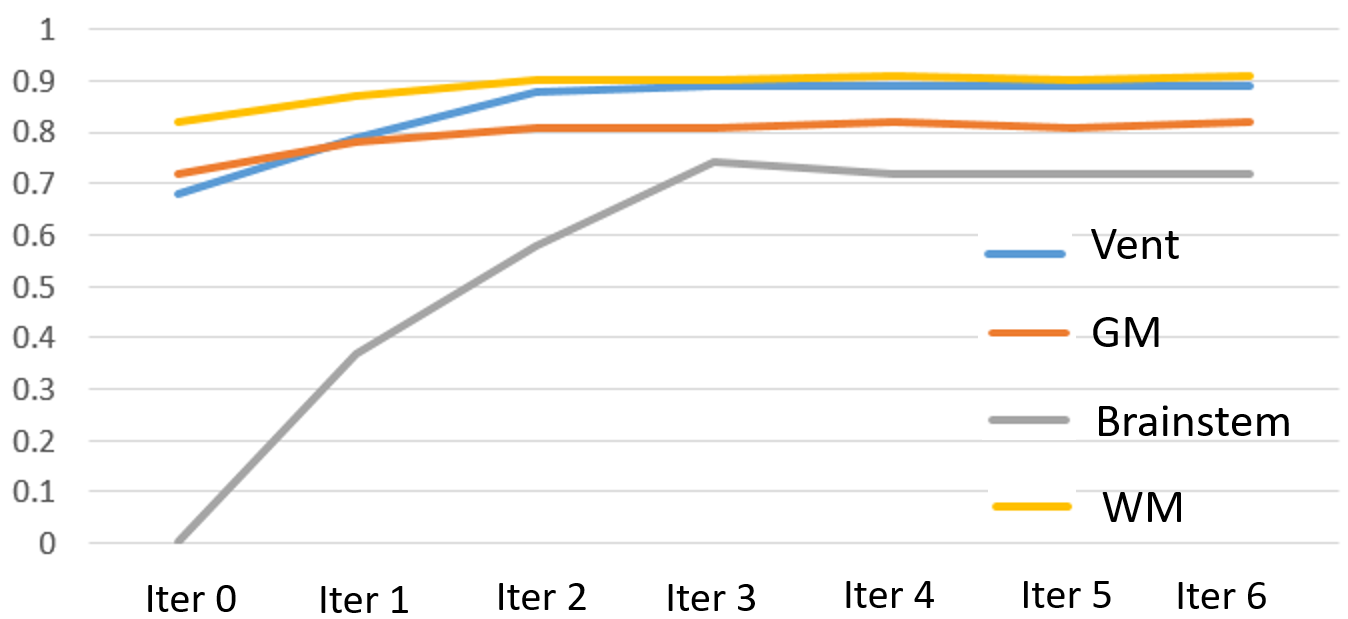} & 
		\includegraphics[height=2.5cm]{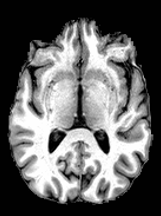} & 			
		\includegraphics[height=2.5cm]{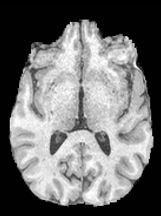}&
		\includegraphics[height=2.5cm]{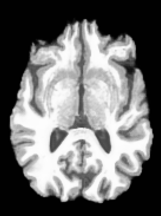}\\
		\footnotesize (a) & \footnotesize (b) & \footnotesize (c) & \footnotesize (d)\\
	\end{tabular}
	\end{minipage}

	\caption{(a) CAMELION Dice coefficients across iterations, (b) Atlas MPRAGE, (b) Corresponding SPGR, (d) Synthetic SPGR.}
	\label{fig:converge_syn}
	\end{center}
	\vspace{-0.4in}
\end{figure}

\begin{wraptable}{r}{2in}
	\vspace{-0.4in}
	\label{tab:correlation}
	\begin{center}
		\scriptsize
		Table 1:  Volume correlations\\
		\begin{tabular}{|l||c|c|c|}
			\hline
			\parbox{0.7in}{Structure} &
			\makebox[0.3in]{SPGR} &
			\makebox[.1in]{NHM} &
			\makebox[.1in]{CAM} 
			\\
			\hline
			Ventricles & 0.859 &  {\bf 0.999} & 0.998 \\
			Gray Matter & 0.875 &  0.902 & {\bf 0.938}\\
			Brainstem & 0.028 &   -0.258 & {\bf 0.794}\\
			White Matter & 0.940 &  0.918 & {\bf 0.971}\\
			\hline
		\end{tabular}
	\end{center}
	\vspace{-0.3in}
\end{wraptable}

Fig.~\ref{fig:converge_syn}(a) shows the convergence of CAMELION over six iterations in terms of the Dice coefficients for a single test data set.  It can be seen that the Dice increases rapidly over the first couple of iterations, and subsequently exhibits only very minor improvements.  Quick convergence is relatively important for this approach, given that the segmentation and synthesis networks need to be re-trained with each iteration.  To reduce training time, each network was initialized with the weights from the previous iteration.  This allowed the number of epochs on the segmentation network to be reduced after the first iteration.  Even so, on a Tesla V100-SXM3 with 32GB of memory, the synthesis training required approximately 20 minutes per iteration, while the segmentation training required approximately 5 hours for the first iteration and 2 hours for subsequent iterations, leading to a total run time of approximately 14 hours (5 iterations).  

Figs.~\ref{fig:converge_syn}(b)-(c) compare the MPRAGE and SPGR images from one of the training data subjects to the synthetic SPGR image computed after running CAMELION.  The atlas images in the training data begins as a pair consisting of the MPRAGE image in Fig.~\ref{fig:converge_syn}(a) and its segmentation.  With each iteration, the synthesis step transforms the MPRAGE image to gain a more SPGR-like appearance.  One notable difference in the accuracy of the synthetic image compared to the original SPGR image is a lack of noise.  Synthesis with U-net architectures have been shown to generally yield rather smooth images~\cite{Dewey19}.

\section{Discussion}

CAMELION provided improved consistency in brain segmentation across two very different T1-weighted imaging protocols. The results shown here were generated essentially with off-the-shelf convolutional neural networks with very limited tuning of the architecture or network parameters.  Furthermore, because there was no gold standard segmentation available on the imaging data set, we employed another segmentation algorithm to generate training data.  This approach, however, had some challenges in that the atlas segmentations lacked sulcal CSF.  We believe that substantial improvements could be gained through careful optimization and the use of improved training data.  

The basic framework of CAMELION bares some similarities to an autoencoder structure.  In the case of CAMELION, the latent space is composed of labels or segmentations that are further transformed into partial volume estimates.  Future work, will investigate the effects of the ``softness'' of the partial volumes, as modulated by the $\beta$ parameter, on the quality of the synthesis.  In addition, it may be possible to bypass the partial volume estimation entirely by directly using the probabilistic results of the segmentation network.  

Although the re-training of networks within each iteration of CAMELION is computationally expensive, when working with data sets with homogeneous acquisition protocols it is possible to  train the network group-wise using all available data.  When a new data set with the same protocol is collected, the segmentation network does not need to be retrained and can be straightforwardly segmented without any additional iterations.  Despite the fact that the data used in this work represents two homogeneously acquired data sets, each image was processed independently to demonstrate that accurate segmentation results can be achieved without training the synthesis network on multiple examples.   A promising area of research will be to use CAMELION in combination with data augmentation and continual learning approaches~\cite{Parisi19} that could lead to increasingly more robust and generalizable segmentation networks.  

\subsection*{Acknowledgements}
This work was supported by a research grant from the National Multiple Sclerosis Society (RG-1907-34570), by the Department of Defense in the Center for Neuroscience and Regenerative Medicine, the intramural research program of the National Institute of Neurological Disorders and Stroke, and the intramural research program of the Clinical Center in the National Institutes of Health.

\newpage
%
%
%
\bibliographystyle{splncs04}
\bibliography{paper}

\end{document}